\documentclass[reprint, superscriptaddress]{revtex4-1}
\usepackage{listings}
\usepackage{graphicx}
\usepackage{dcolumn}
\usepackage{upgreek}
\usepackage{bm,soul}
\usepackage{braket,algorithm}
\usepackage{algorithmic}
\usepackage[colorlinks,linkcolor=blue,anchorcolor=blue,citecolor=blue,urlcolor=blue]{hyperref}
\usepackage{tikz}
\usepackage{booktabs,multirow,mathrsfs,amssymb,amsmath,amsthm}
\usepackage{caption}
\usepackage{makecell}
\usepackage{qcircuit}
\usepackage{subfig}
\usepackage{float}

\newtheorem{definition}{Definition}
\newtheorem{theorem}{Theorem}

\begin{document}

\title{Quantum-Classical Hybrid Algorithm for Solving the Learning-With-Errors Problem  on NISQ Devices}

\author{Muxi Zheng}
\affiliation{Beijing Academy of Quantum Information Sciences, Beijing 100193, China}
\affiliation{State Key Laboratory of Low-Dimensional Quantum Physics and Department of Physics, Tsinghua University, Beijing 100084, China}

\author{Jinfeng Zeng}
\affiliation{Beijing Academy of Quantum Information Sciences, Beijing 100193, China}

\author{Wentao Yang}
\affiliation{State Key Laboratory of Low-Dimensional Quantum Physics and Department of Physics, Tsinghua University, Beijing 100084, China}

\author{Pei-Jie Chang}
\affiliation{State Key Laboratory of Low-Dimensional Quantum Physics and Department of Physics, Tsinghua University, Beijing 100084, China}

\author{Quanfeng Lu}
\affiliation{State Key Laboratory of Low-Dimensional Quantum Physics and Department of Physics, Tsinghua University, Beijing 100084, China}

\author{Bao Yan}
\affiliation{State Key Laboratory of Mathematical Engineering and Advanced Computing, Zhengzhou 450001, China}

\author{Haoran Zhang}
\affiliation{Division of Physics and Applied Physics, School of Physical and Mathematical Sciences, Nanyang Technological University,  Singapore 637371, Singapore}

\author{Min Wang}
\affiliation{Beijing Academy of Quantum Information Sciences, Beijing 100193, China}

\author{Shijie Wei}
\email{weisj@baqis.ac.cn}
\affiliation{Beijing Academy of Quantum Information Sciences, Beijing 100193, China}

\author{Gui-Lu Long}
\email{gllong@tsinghua.edu.cn}
\affiliation{Beijing Academy of Quantum Information Sciences, Beijing 100193, China}
\affiliation{State Key Laboratory of Low-Dimensional Quantum Physics and Department of Physics, Tsinghua University, Beijing 100084, China}
\affiliation{Frontier Science Center for Quantum Information, Beijing 100084, China}
\affiliation{Beijing National Research Center for Information Science and Technology, Beijing 100084, China}

\begin{abstract}
	The Learning-With-Errors (LWE) problem is a fundamental computational challenge with implications for post-quantum cryptography and computational learning theory. 
	 Here we propose a quantum-classical hybrid algorithm with Ising model   to address LWE, transforming it into the Shortest Vector Problem and using variable qubits to encode lattice vectors into an Ising Hamiltonian. By identifying low-energy Hamiltonian levels, the solution is extracted, making the method suitable for noisy intermediate-scale quantum  devices. The required number of qubits is less than $m(m+1)$, where $m$ is the number of samples. Our heuristic algorithm's time complexity depends on the specific quantum eigensolver used to find low-energy levels,   and the performance when using the Quantum Approximate Optimization Algorithm is investigated.  We validate the algorithm by solving a $2$-dimensional LWE problem on a $5$-qubit quantum device, demonstrating its potential for solving meaningful LWE instances on near-term quantum devices.
\end{abstract}

\maketitle

\section{Introduction}\label{Sec1}

Shor's algorithm \cite{shor}, one of the most important quantum algorithms, is able  to  factor large integers in polynomial time, posing a significant threat to RSA \cite{RSA}.
To address this challenge,  post-quantum cryptography \cite{bernstein2017post} has been proposed,  which aims to provide safe encryption that are resistant to attacks from both classical and quantum computers.
Among various post-quantum encryption protocols, algorithms based on the Learning With Errors (LWE) problem \cite{lwe} have attracted the biggest attention.  
The hardness of the LWE problem is thought to be equivalent to solving the worst-case lattice problems \cite{hardness}, which are considered computationally difficult and  impossible to solve in polynomial time even by quantum computers.

The best known classical algorithm for solving the LWE problem is of subexponential time complexity \cite{blum2003noise,arora2011new}.   Several quantum algorithms have been proposed to tackle the LWE problem. When quantum samples are well-prepared,  algorithms\cite{quantumsample1, quantumsample2} can be highly efficient. However, the preparation of quantum samples presents significant challenges. Quantum walks have been employed to address the ternary-LWE problem, a specific variant of the LWE problem \cite{quantumwalk}. Additionally, a quantum search-based algorithm has been proposed for the general LWE problem \cite{quantumsearch}. Despite their potential, these algorithms suffer from exponential increases in circuit depth as the problem size grows, making them impractical for implementation on near-term quantum devices.
    To address this challenge, the variational quantum algorithm is implemented after transferring the LWE problem into bounded distance decoding  (BDD) problem. \cite{lv2022using}

There are two versions of the LWE problem: LWE-decision problem and LWE-search problem, which have been proven to be equivalent \cite{lwe, Albrecht2015}.  We will focus specifically on the  LWE-decision problem in this paper.  
The LWE-decision problem can be transformed into the Short Integer Solution (SIS) problem \cite{Albrecht2015, SIS}, which can be solved by finding the short enough vector in the lattice.   The time complexity of the transforming process is polynomial,  consequently, the most time-consuming part of the LWE-decision problem is to find the short enough vectors in a lattice, which requires exponential time to calculate. For the  shortest vector problem (SVP) in a lattice, several related  works are proposed to  use adiabatic quantum computation (AQC) \cite{joseph2020not, joseph2021two,yamaguchi2022annealing}. The  methods are extended to variational quantum algorithm (VQA) subsequently \cite{Albrecht2023}.

In this paper, we propose  a quantum-classical hybrid algorithm with Ising model (HAWI) to solve the LWE-decision problem. After a series of classical preprocessing via the SIS problem, we construct the problem Hamiltonian of the LWE-decision problem. In the algorithm, we map each eigenstate of the Hamiltonian to a corresponding vector in the lattice, and its eigenvalue is equal to the norm of the vector. By finding the low-energy levels of the given Hamiltonian, one can obtain short vectors in the lattice. If short enough vectors are encompassed in eigenvectors of the Hamiltonian, one can obtain the solution to the SIS problem. After classical postprocessing, the LWE-decision problem is solved. We prove that the qubit number required for our algorithm is polynomial with the problem size, rendering it friendly for implementation in real quantum devices.
The running time of our algorithm depends on the time complexity of the algorithm for finding the low-energy state of the Ising Hamiltonian.  We 
focus on QAOA and conclude
that  if the number of iterations to a quantum state  which corresponds to  success probability $P_r$ satisfies $y < O\left(P_r\cdot m\log m\cdot 2^{0.2972k}/pk^2\right)$,  
 our method will exhibit advantages over classical BKZ algorithms, 
 where $k$ is the block size related to the problem parameters and $p$ is the number of layers in QAOA.  We  propose a heuristic parameter design that can improve the success rate of QAOA for LWE problems.  Finally, we implement the algorithm in the IBM quantum platform to demonstrate the feasibility of our algorithm  on the NISQ devices.

\section{Results}
\subsection{The framework of the algorithm}\label{algo}

The LWE-decision problem is the problem of deciding whether pairs $(\mathbf{a},c)\in \mathbb{Z}^n_q\times\mathbb{Z}_q$ are sampled according to a specific probability distribution $L_{s,\chi}$ or the uniform distribution on $\mathbb{Z}^n_q\times\mathbb{Z}_q$. The LWE-decision problem is specified by parameters $n, q, \sigma$ and $m$, where $n$ is the parameter determining the problem size, $m$ is the number of samples given in the problem,  
and $\sigma$ is related to the variance of the probability distribution $L_{s,\chi}$.   Rigorous definition and detailed explanation  
    is shown in `Methods'. 

\begin{figure*}
	\centering
	\includegraphics[width=0.9\textwidth]{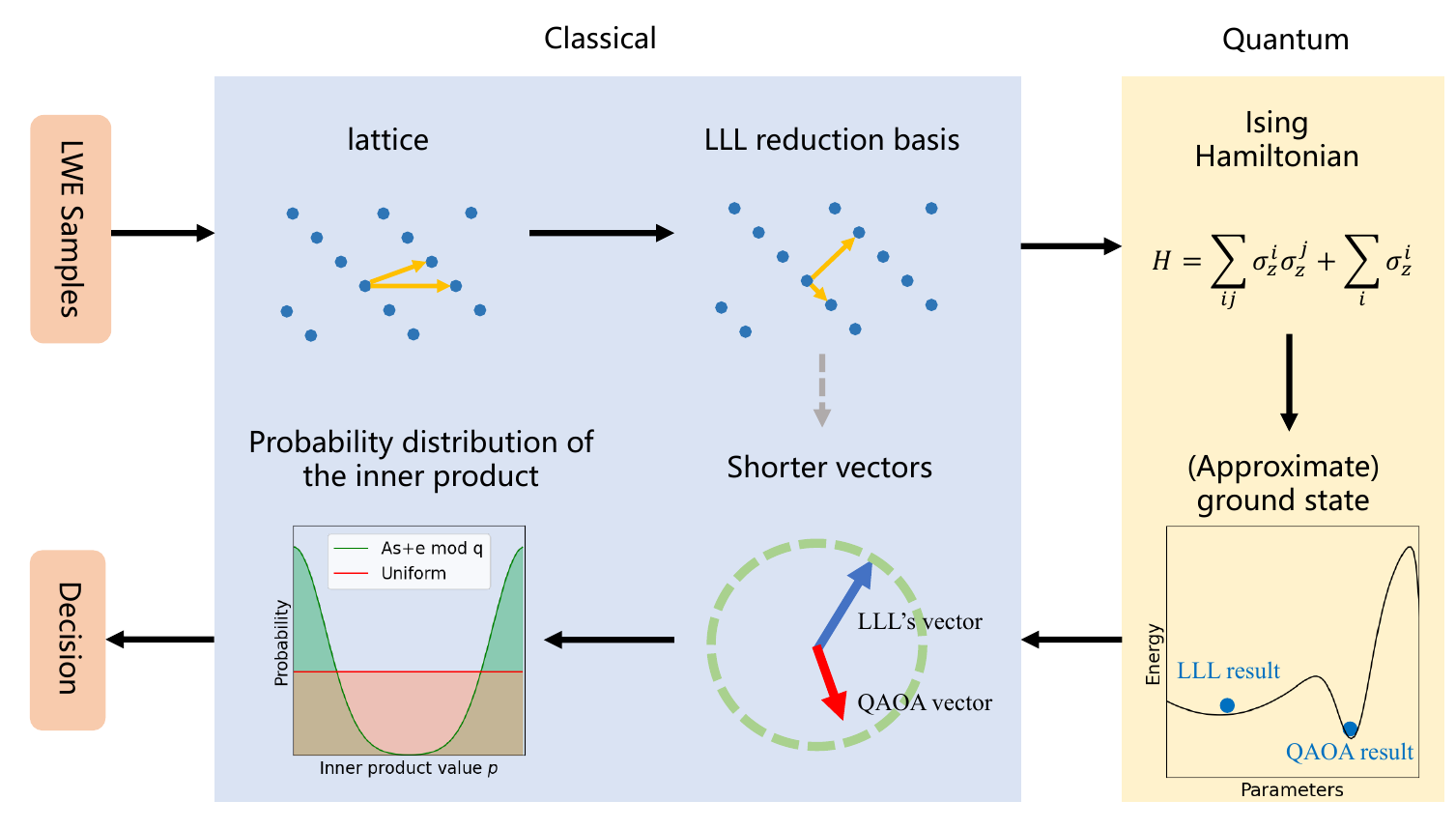}
	\caption{   The workflow of the HAWI algorithm for the LWE-decision problem.  Firstly, we use classical techniques to transfer the LWE samples into LLL reduction basis. Secondly, we construct the Ising Hamiltonian of the LWE problem and utilize quantum optimization algorithm (such as QAOA) to find the shorter vector  which is closer to the solution. Finally, we use the shorter vector to calculate the inner product of the vectors and determine which distribution it satisfies to output the decision result of the LWE problem. 
	}\label{workflow}
\end{figure*}

 We  induce quantum optimization to  classical LWE-decision algorithm to seek possible acceleration. The workflow of our algorithm is shown in Fig. \ref{workflow}. 
 In the following, we introduce the framework of the algorithm in brief. 

 Classically,  solving the LWE-decision problem can be transferred into  
finding the sufficiently short vectors in the lattice\cite{Albrecht2015,Bindel}.  
From the samples $(\mathbf{a}_i, c_i)$ provided in the problem, we make calculation to obtain a set of vectors $\mathbf{b}_1, \mathbf{b}_2, ... , \mathbf{b}_m$, and define a lattice $\mathcal{L}=\{\mathbf{v}\ | \ \mathbf{v}=\sum_i^my_i\mathbf{b}_i\}$. The next step is to find the sufficiently short vectors in the lattice.   
After the short vectors are obtained, we calculate the inner products $I_p$ as a function of the vectors. The procedure is repeated  to determine which distribution $I_p$ follows. From the distribution, we output the decision. The procedure of this approach is described in detail in `Methods' In the procedure, giving  a constant distinguish advantage $\epsilon$,  finding the short enough vector in the lattice with the length
\begin{equation}\label{length}
	||\mathbf{v}||\le\frac{q}{\sigma\uppi}\sqrt{\frac{1}{2}\ln\frac{1}{\epsilon}},
\end{equation}
is the most time-consuming part. 
There are several classical algorithms to make basis reduction and obtain the short vectors in the lattice, such as LLL algorithm and BKZ algorithm (See `Methods' for details).
Generally, the algorithms that have polynomial time complexity can't  guarantee enough accuracy, while the algorithms which can find 
 short enough vectors will always exhibit exponential time complexity.

Here, we utilize a quantum algorithm \cite{joseph2021two, Albrecht2023} to find a short enough vector. From the LLL reduction basis $\mathbf{b}_1, \mathbf{b}_2, ... , \mathbf{b}_{m}$, we construct the Hamiltonian 
\begin{equation}\label{eq4}
	H = ||\sum_{i=1}^{m}\hat{x}_i\mathbf{b}_i||^2
	=\sum_{j=1}^m\big(\sum_{i=1}^{m}\hat{x}_ib_{i,j}\big)^2,
\end{equation}
where $b_{i,j}$ represents $j$-th component of vector $\mathbf{b}_i$. We encode $\hat{x}_i$  on $\xi_i$ qubits using Pauli matrices according to the following rules \cite{joseph2021two,sqif}
\begin{equation}
 \hat{x}_i = \sum_{j=1}^{\xi_i} 2^{j-2}\sigma^z_{i,j}-\frac{1}{2},
\end{equation}
where $\sigma^z_{i,j}$ represents the 
 Pauli-Z matrix acts on the $j$-th encoding qubit of $\hat{x}_i$,  abbreviated from $I^{\otimes(j-1)}\otimes\sigma^z\otimes I^{\otimes(\xi_i-j)}$. 
Eq. (\ref{eq4}) is the familiar Ising model Hamiltonian.

We denote the  eigenstate of $\hat{x}_i$ as $\ket{x_i}$ with corresponding eigenvalue $x_i$. 
Then the eigenstate $\ket{\lambda_j}$  of the Hamiltonian  can be represented as $\ket{\lambda_j} = \ket{x_1}\otimes\ket{x_2}\otimes ... \otimes \ket{x_{m}}$. 
  Each  $\ket{\lambda_j}$ corresponds a vector $\mathbf{v}_j = \sum_{i=1}^{m}x_i\mathbf{b}_i$ in the lattice, and the energy of the eigenstate $\ket{\lambda_j}$ is equal to  $|| \mathbf{v}_j ||^2$.  Hence, a shorter vector corresponds to a lower energy level. 
   As long as the  energy of the eigenstate satisfies the Eq. (\ref{length}) for certain $\epsilon$, the corresponding eigenstate $\mathbf{v}$  can be chosen as a candidate  vector to  solve the LWE-decision problem.  {Since shorter vector $\mathbf{v}$ induces a larger distinguishing advantage $\epsilon$,} the largest distinguishing advantage $\epsilon_{\max}$ is achieved in the  shortest non-zero vector, which corresponds to the first excited state of the Hamiltonian.
To find the low energy levels of the Ising Hamiltonian (\ref{eq4}), several methods, such as QAOA \cite{farhi2014quantum}, full quantum eigensolver (FQE) \cite{wei2020full}  and quantum annealing \cite{finnila1994quantum}, can be utilized.

In our HAWI algorithm, the number of qubits required is $\sum_{i=1}^m\xi_i$. This parameter is closely related to the success probability of  obtaining a vector of a short enough length. A larger value of $\xi_i$ corresponds to a higher  probability of success in this regard.  In the next subsection `Complexity Analysis', we will provide a theoretical upper bound for the number of qubits, ensuring that the shortest vector in the lattice will be certainly included in the eigenstate of the Hamiltonian.

Alternatively, we can construct the Hamiltonian in the following way \cite{yamaguchi2022annealing}
\begin{equation}\label{alter_ham}
	H_r = \sum_{j=1}^m(\sum_{i=1}^{r-1}\hat{x}_ib_{i,j}+\hat{y}_rb_{r,j})^2,
\end{equation}
where $r=1,2,...,m$ and 
\begin{equation}
	\hat{y}_r = \sum_{j=1}^{\xi_r}2^{j-2}(\sigma^z_{r,j}+1)+1.
\end{equation}
Denoting the ground state and ground energy of $H_r$ as $\ket{g_r}$ and $E_r$, respectively,
state $\ket{g_r}$ with the smallest $E_r$ among $E_1, E_2, ... , E_m$ corresponds to the shortest non-zero vector in the lattice.
The advantage of this Hamiltonian encoding scheme is that it avoids the ground state of Hamiltonian being a zero-norm vector in the lattice.

\subsection{Complexity Analysis}\label{Com}

We  derive a theoritical upper bound for the number of qubits based on the property of LLL basis. The comprehensive proof of this theoritical upper bound  is provided in `Supplementary  Note 2'. Here, we present the  conclusions.

\begin{theorem}\label{theo1}
	{If we use}
	\begin{equation}
		\xi_{m-i} = \log_2\Big(
		\alpha_i\cdot(\delta-\frac{1}{4})^{\frac{1-m}{2}}\Big)
	\end{equation}
	{qubits to encode $(m-i)$-th LLL basis, where $i=0,1, ... , m-1$, and $\alpha_i$ satisfies the following recursive equation}
	\begin{equation}
		\alpha_i = (\delta-\frac{1}{4})^{\frac{i}{2}} + \frac{1}{2}\sum_{k=0}^{i-1}\alpha_k,
	\end{equation}
	{with initial condition $\alpha_0=1$, then the Hamiltonian in Eq. (\ref{eq4}) is capable of including the shortest non-zero vector in the lattice,
		where
		$\delta\in(1/4, 1)$ is the parameter in the LLL algorithm.}
\end{theorem}

Specifically, by setting $\delta = \frac{3}{4}$ in Theorem \ref{theo1}, we obtain $\xi_{m-i} \leq \lceil \frac{m +1}{2}\rceil + i$. This leads to the total number of qubits to
\begin{equation}
	N = \sum_{i=0}^{m-1}\xi_{m-i} \le m(m+1). 
\end{equation}
Consequently, by encoding $N_{\max}=m(m+1)$ qubits in the  Hamiltonian, we ensure  the inclusion of the shortest vector in the  lattice. We notice that there are related works\cite{Albrecht2023,kannan1983improved} regarding the required qubit number in SVP problems, which are introduced in `Supplementary Note 4'. 

In practice, LLL reduction usually generates much shorter vectors compared to its theoretical bounds, leading to the required qubit number for SVP being practically much fewer than its theoretical bound $N_{\max}$. Therefore, we can just run the algorithm based on the LLL basis, utilizing fewer qubits than its theoretical bound in practice to save quantum resources.   In the next subsection `Performance of HAWI', we will show the necessary qubit number required by numerical simulation for the small-size LWE-decision problem.

The time complexity of our algorithm  can be represented as
	$T = t_G + t_L + t_q/\epsilon^2$,
where $t_G=O(m^2n)$ is the running time of the Gaussian elimination method, and  $t_L$ is the running time of the classical LLL algorithm \cite{LLL} , which is implemented in Step 2 of Algorithm \ref{lwe_alg}.  $t_q$ represents the runtime for finding the low-energy eigenstates that satisfy the condition  in Eq. (\ref{length}) for a given Hamiltonian. This runtime depends on  the quantum optimization method we choose. Since $t_G$ and $t_L$ increase polynomially with problem size, the time complexity $T$ is dominated by $t_q$. 

We consider the time usage of the quantum algorithm in comparison of the classical algorithm. For QAOA, the time complexity can be expressed as $t_q =y\cdot pO(m^2)/P_r$, where $O(m^2)$ denotes the time complexity of running  each layer of the QAOA circuit,  $p$ represents the number of layers,  $y$ represents the number of iterations needed to evolve the initial state to low-energy states,  and $P_r$ represents the overlap between the low-energy state and the states corresponding to the sufficiently short vectors, which is exactly the success probability of the QAOA.
The  value of $y$ and $P_r$ for given parameters is unknown, leading to the time complexity of our algorithm unclear. However, we can make some heuristic comparison. Under the following conditions
\begin{equation}
	y <O\Big(\frac{m\log m{P_r}}{pk^2}\cdot 2^{0.2972k}\Big),
\end{equation}
 the time complexity $t_q$ of quantum algorithm will be shorter than the BKZ algorithm, namely, our algorithm will exhibit the advantage compared to the classical BKZ algorithm. The block size $k$ chosen in the BKZ algorithm is determined by the parameters in the LWE-decision problem in the following way: $k={m}/{\log \big(q^{1-n/m}/\sigma\uppi\cdot \sqrt{\ln(1/\epsilon)/2}\big)}$. (See `Methods' for details.)

In the Table \ref{table1},  we summarize the quantum resources consumed of  our quantum algorithm, and  compare them with those from the classical BKZ algorithm.

\subsection{Performance of HAWI}\label{simu}

 We demonstrate the workflow of the LWE-decision algorithm by numerical simulation, and present a numerical analysis regarding the suitable number of qubits and the performance of the algorithm. We denote the shortest vector in the LLL basis as $\mathbf{b}_0$, and the shortest vector in the lattice as $\mathbf{v}_0$.  In this subsection, we don't care which quantum optimization algorithm utilized to find the ground state, and assume that we will obtain the vectors as long as they are encoded by the Hamiltonian. We will specifically focus on QAOA and study the success probability to find the ground state in the next subsection.

Firstly, we demonstrate how the LWE-decision algorithm works. 
The vector $\mathbf{s}$ is randomly generated in $\mathbb{Z}^n_q$.   Half of the samples are sampled according to $L_{s,\chi}$, others are uniform distribution on $\mathbb{Z}^n_q\times \mathbb{Z}_q$. We follow the procedure described in Algorithm \ref{lwe_alg} and obtain the probability distribution of inner product $I_p = \braket{\mathbf{v},\mathbf{c}}_q$ of them separately, which is shown in Fig. \ref{simulation}(a).   It can be observed that for the instances sampled according to $L_{s,\chi}$,   $I_p$  follows the   Gaussian distribution $\chi$, and for those instances sampled randomly, $I_p$ follows uniform distribution. See   `Supplementary Note 5' for more details of  the simulation.

Then we demonstrate the number of qubits  required of our algorithm for small sized problems.
For the lattices generated by the LWE-decision problem, we solve the SVP and obtain the shortest vector $\mathbf{v}_0$, which can be written as $\mathbf{v}_0 = \sum_{i}^{m}x_i\mathbf{b}_i$ using LLL basis $\mathbf{b}_i$. By determining the maximum value of coefficient $|x|_{\text{max}}=\max\{|x_1|, |x_2|, ... , |x_m|\}$, we can calculate the number of qubits required $n_e$ per basis for each instance by the equation $n_e =\lceil \log_2 |x|_{\max} \rceil+1$.
The simulation results and  the comparison  with the BKZ algorithm is shown in Fig. \ref{simulation}(b). 
The parameter is set as $n=10$, $m=30$, $q=101$.
 In the simulation, we neglect the instances that $\mathbf{b}_0$ is already the shortest vector in the lattice, and implement both the BKZ algorithm and our quantum approach to find shorter vectors for the rest instances. We observe the successful probability  above 97\% of  finding the shortest vector of the lattice when we encoding each LLL basis with 3 qubits. This is better than the results obtained by the BKZ algorithm for $k<12$. 
This indicates that $3m=90$ qubits are enough in this case to generate good enough result, which is much smaller than the theoretical upper bound $(m+1)m=930$ for $m=30$.

Subsequently, we demonstrate how the result changes as problem size $n$ increases.  The parameters change with $n$ in the following ways, $m=2n$, $q\approx n^{2}, \sigma = \sqrt{2n/\uppi}$. The upper figure in Fig. \ref{simulation}(c) shows that the proportion of instances that $\mathbf{v}_0$ is shorter than $\mathbf{b}_0$ as problem size $n$ increases,  which indicates that the LLL basis will be gradually far away from the shortest vector in the lattice for larger lattice dimension. The lower figure in Fig. \ref{simulation}(c) shows that the success probability of finding the shortest vectors when we use $z$-qubits encoding each LLL basis. As problem size $n$ increases, the value of $z$ should increase to maintain a constant success probability.
Once again, we find that the required  number of qubits in practice is much smaller than the theoretical upper bound for small-sized lattice.

\subsection{Numerical and experimental results of QAOA}\label{experiment}

 We simulate the algorithm using QAOA and analyze the success probability of obtaining the ground state. In our simulation, we set $m=2n$ and select instances where one qubit per LLL basis is sufficient to identify the shortest vector. Such instances are not uncommon at this scale of the problem in our simulations. The success probability of QAOA in our algorithm is primarily influenced by the number of qubits. Therefore, the success probability results obtained from our simulation remain applicable even for instances where more qubits are required to encode each LLL basis.

We use the Hamiltonian encoding method according to Eq. (\ref{alter_ham}) to eliminate the eigenstate with eigenvalue $E=0$. The Hamiltonian is constructed as
 $H_i=(\mathbf{b}_i - \sum_{j\ne i}x_j\mathbf{b}_j)^2$, where $x_j=(1+\sigma_j^z)/2$.  Using the algorithm we  described in  subsection `The framework of the algorithm' and section  `Method', we can obtain the LLL reduction basis $\{\mathbf{b}_1, \mathbf{b}_2, ... , \mathbf{b}_{m}\}$ and the corresponding Hamiltonian $H\in\{H_1, H_2, ... , H_{m}\}$. The QAOA circuit can be written as the following unitary operator
 \begin{equation}
 	U(\mathbf{\beta}, \mathbf{\gamma}) = \prod_{k=1}^{p}e^{i\beta_k\sum_j\sigma^x_j}\cdot e^{-i\gamma_k H},
 \end{equation}
where $\beta, \gamma$ are the parameters that need to be  optimized. 
We take the expectation value of Hamiltonian $E(\mathbf{\beta}, \mathbf{\gamma}) = \braket{\phi_0|U^\dagger((\mathbf{\beta}, \mathbf{\gamma})) H U(\beta,\gamma)|\phi_0}$ as the cost function, where the initial state is $\ket{\phi_0}=\ket{+}^{\otimes n}$.

\begin{figure*}
	\centering
	\includegraphics[width=0.95\textwidth]{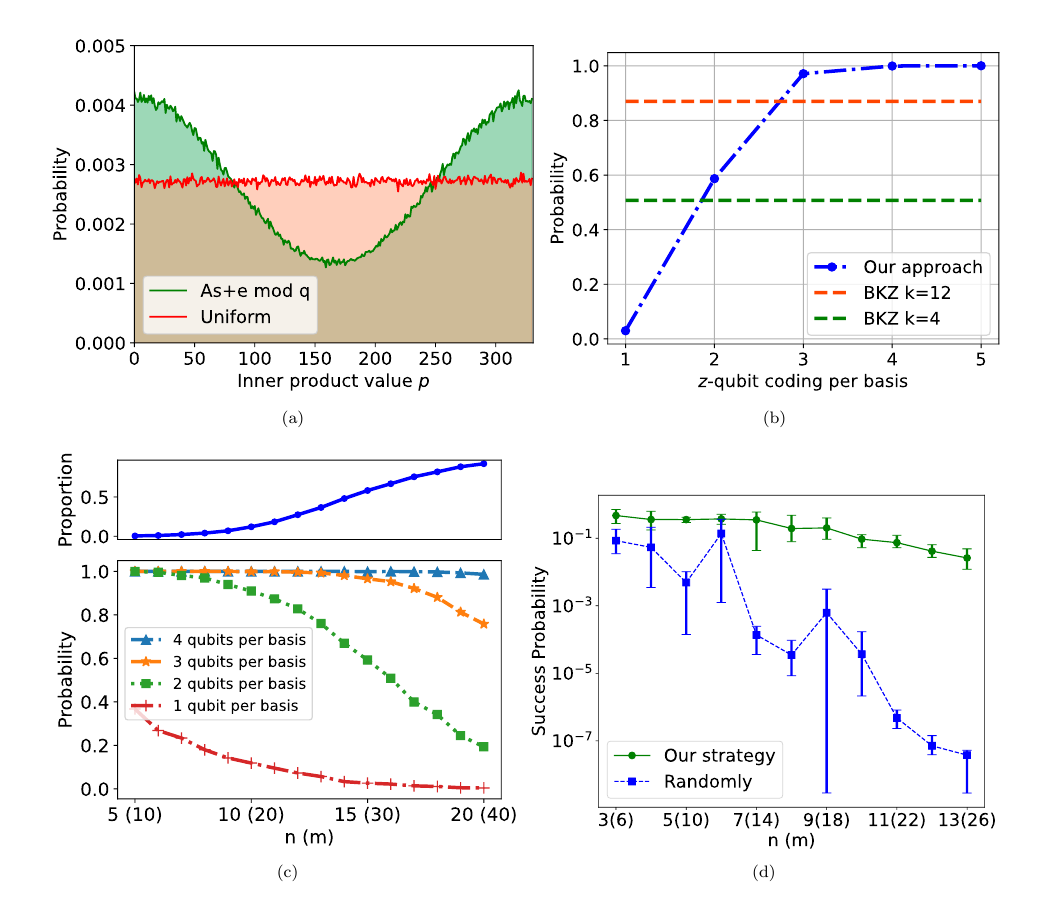}
	\caption{  Numerical results regarding the performance of our algorithm. (a) Probability distribution of inner product $I_p = \braket{\mathbf{v}, \mathbf{c}}_q$ on $M$ instances. 
		Green  line represents the results from the instances sampled according to $L_{s,\chi}$, while the red line represents the results from the instances following uniform distribution.
		Parameters: $n=18$, $m=36$, $\sigma=3$, $q=331$.
		(b)
		The success probability of finding the shortest vector in the lattice when we use $z$ qubits to encode each LLL basis. In the simulation process, we discard the instances that  the shortest vector in the lattice is already contained in the LLL basis.
		We compare this result with the BKZ algorithm with different block $k$. 
		Parameters: $n=15$, $m=30$, $q=101$.
		(c) Upper:
		With the growth of the problem size $n$,
		the proportion of  instances that $\mathbf{v}_0$ is shorter than $\mathbf{b}_0$.
		Lower: With the growth of $n$ ($m$), 
		 the success probability of finding the shortest vector in the lattice when we use $z$ qubits to encode each LLL basis. 
		(d) The success probability of QAOA with different problem size $n$. The green line represents the result by utilizing our heuristic strategy for parameters initialization, while the dotted blue line represents the results by randomly choosing the initial parameters. The number of qubits is $n_{\text{q}} = m-1 = 2n-1$, and layers of QAOA are chosen as $p=n_{\text{q}}$.  The error bar on each data point $(n_i,P_i)$ extends from $(n_i,P_{i}^{\min} )$ to $(n_i, P_i^{\max})$, where $P_i^{\min}$ and $P_i^{\max}$ are the minimum and maximum value among $R$ simulation results $P_{ij}$, $j=1,2, ... , R$ for each point.}\label{simulation}
\end{figure*}

 The performance of the QAOA is closely related to the choice of initial parameters \cite{zhou2020quantum}. 
Therefore, rather than randomly generating them, we try to find an efficient strategy for choosing the initial parameters. Under the observation of the patterns for optimal parameters, we propose a heuristic formula for the  initial parameters. We describe the observed patterns and give an explanation why this choice may be efficient in `Method'. We only show the conclusion here. We choose the initial parameters as follows
     \begin{equation}\label{eq11}
     	\gamma_k = -\frac{2\uppi}{c}\cdot\frac{k}{p+1},\qquad
     	\beta_k = \frac{\uppi}{4}\cdot\frac{p-k+1}{p},
     \end{equation}
where $c$ is the constant term in the Hamiltonian $H = \sum_{ij}a_{ij}\sigma^z_i\sigma^z_j+\sum_ib_i\sigma^z_i+c$ by expanding Eq. \ref{eq4} or \ref{alter_ham}. 
     We simulate the QAOA to study the success probability under the relation $p=n_{\text{q}}$, where $n_{\text{q}}$ is the number of the qubits. The results are shown in Fig. \ref{simulation}(d). We can see that the success probability by following Eq. \ref{eq11} is significantly higher than that obtained by choosing the initial parameters randomly, which demonstrates that the efficiency of the QAOA is satisfactory with our setting for the initial parameters.  More detailed performance analysis of the QAOA optimization  is provided in `Supplementary Note 7'.

\begin{figure*}
	\centering
	\includegraphics[width=\textwidth]{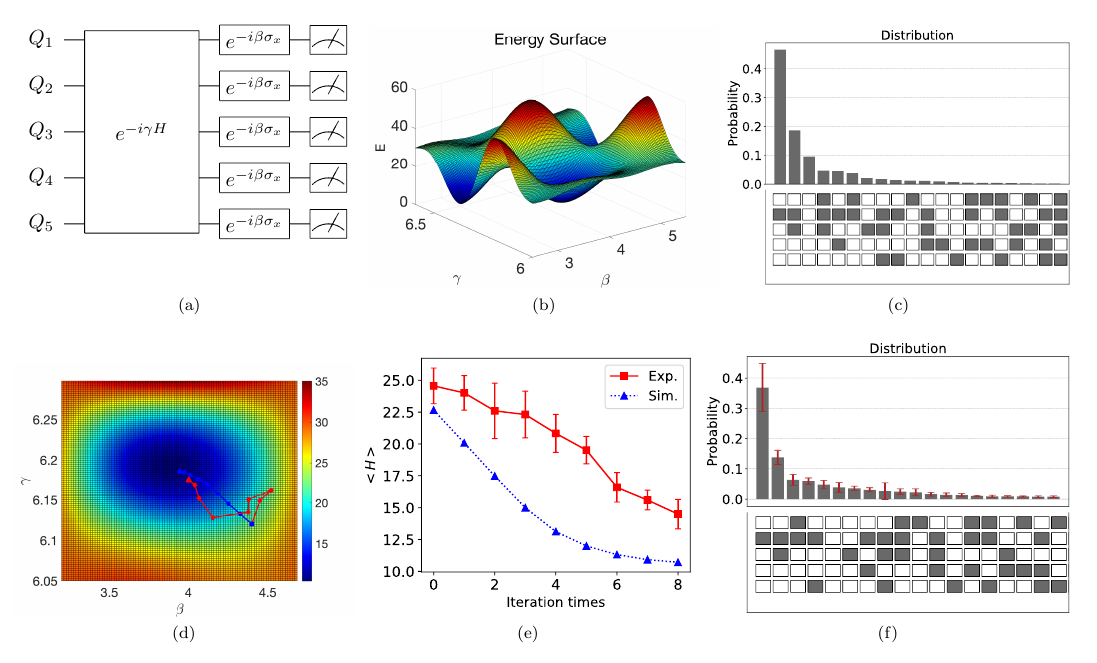}
	\caption{Numerical simulation and experimental results. (a) Quantum circuit of HAWI-QAOA. 
		(b) The energy surface formed by  $E(\beta, \gamma)=\braket{\phi_0|U^\dagger(\beta,\gamma) HU(\beta,\gamma)|\phi_0}$. 
		(c) Probability distribution of each computational basis, $\beta=3.88$, $\gamma=6.19$, where Hamiltonian reaches minimum. The white block represents 0 and  colored block represents 1 in $x$-tick labels. $\ket{01000}$ corresponds the shortest vector in the lattice.
		(d) Optimization process shown in parameter space, corresponding to that in (b).  The red curve and the blue curve represent experimental results and numerical simulation results respectively.
		The initial parameters are  $\beta=4.40, \gamma=6.12$ with  expected $E=25.44$. After iterations,  $E=14.15$ with $\beta=4.00, \gamma=6.18$. 
		(e) Expectation value of Hamiltonian for each iteration.  The red curve and the blue curve represent experimental results and numerical simulation results respectively. 
		(f) The experimental results of probability distribution of each computation basis. The probability of the target state is 0.37.
		The error bar in (e) and (f) on each data point $(x_i, P_i)$ extends from $(x_i, P_i - \sigma_i)$ to $(x_i, P_i + \sigma_i)$, where $\sigma_i =\sqrt{ \frac{1}{R} \sum_{j=1}^R(P_{ij}- \frac{1}{R} \sum_{k=1}^RP_{ik})^2}$ for $R$ repeated measurement results $P_{ij}$, $j=1,2, ... , R$.}\label{instance}
\end{figure*}

We demonstrate an example on a real quantum device using QAOA to solve the LWE-decision problem. We set the parameters as $n=2$, $m=6$, $q=17$. 
Therefore, 5 qubits are required. For single-layer QAOA, the unitary operation can be expressed as
\begin{equation}
	U(\beta, \gamma) = e^{-i\beta\sum\limits_i\sigma_i^x}\cdot e^{-i\gamma H}.
\end{equation}
The quantum circuit to implement $U(\beta,\gamma)$ is shown in Fig. \ref{instance}(a), 
    which is feasible to be implemented in the present superconducting quantum devices. The value of $E(\beta, \gamma)$ around the optimal point $(\beta_0,\gamma_0)$ is illustrated in Fig. \ref{instance}(b). The  probability distribution of each computational basis at point $(\beta_0, \gamma_0)$ is shown in Fig. \ref{instance}(c). The appearance probability of the ground state is $P=0.47$, which is  fifteen times larger than the average probability of $1/2^5\approx0.03$, showing the significant enhancement of a single-layer QAOA optimization.

We conduct the  experiment on the IBM quantum platform \cite{kyoto}, using the instance above. 
We  obtain the expectation value of $E(\beta,\gamma)$ for single-layer QAOA, and use the gradient descent algorithm for parameter optimization (See `Supplementary Note 6' for details).   The experimental results in the parameter space are shown in Fig. \ref{instance}(d), with the parameters optimized in the direction of the red lines. After 8 iterations, we achieve a relatively small expectation value of Hamiltonian. For comparison, we also present the iteration results of the numerical simulations (blue line). The expectation value of Hamiltonian at each iteration in the experiment and the numerical simulation is illustrated in Fig. \ref{instance}(e).  From the experimental probability distribution  shown in Fig. \ref{instance}(f), we conclude that  the probability of the target state is improved to $0.37$ after the optimization, which is close to the simulation results.

\renewcommand\arraystretch{1.5}
\begin{table*}
 \caption{Resources required for LWE-decision problems with classical and quantum algorithms. }\label{table1}
 \resizebox{1.0\textwidth}{!}{
  \begin{tabular}{cccc}
   \toprule[1.5pt]
   & BKZ algorithm (using sieving) & BKZ algorithm (using enumeration)&HAWI with QAOA  \\
   \midrule[1pt]
   Space  Complexity & $2^{0.2972k}$ & $\text{poly}(k)$ &$O(m^2)$\\
   Time Complexity &$\frac{m^3}{k^2} \cdot 2^{0.2972k}\log m/\epsilon^2$ & $\frac{m^3}{k^2}\cdot 2^{O(k^2)}/\epsilon^2$ &$yO(m^2)\cdot P_r/\epsilon^2$  \\
   \bottomrule[1.5pt]
   \end{tabular} }
\end{table*}

\section{Discussion} \label{conclu}

In this paper, we introduce a quantum classical hybrid algorithm with Ising model, HAWI, to solve the LWE problem.  We provide an upper bound for the number of qubits that ensures the inclusion of the solution in the Hamiltonian of LWE problem. 
The runtime of our algorithm, which depends on the approaches for finding the approximate ground state of Hamiltonian, is also discussed. Although the exact conclusion is not clear in general, some heuristic results can be obtained.
For QAOA, in the circumstances that the number of  iterations  $y < O\left(P_r m\log m\cdot 2^{0.2972k}/(pk^2)\right)$, the HAWI  algorithm will exhibit advantages over classical algorithms.
Furthermore, we  experimentally demonstrated our algorithm on a superconducting hardware with 5 qubits. 

To the best of our knowledge, our work represents the first attempt to utilize quantum hardware to address small-scale LWE problems, providing valuable insights and experience for tackling larger-scale problems in the future. The analysis of the upper bound on quantum qubit resources establishes a theoretical limit on the resources required for lattice-based quantum algorithm design. Additionally, the advancements in the design of QAOA parameters contribute to improving the success rate of such algorithms when applied to LWE problems. Our algorithm is heuristic, so the performance  for larger problem size is worthwhile to be studied in the future.

Currently, our method does not pose an immediate threat to post-quantum cryptographic systems. The computational complexity of QAOA remains unclear, and its success rate may decrease exponentially with problem size.  While it is not impossible that this approach could lead to subexponential complexity, it is more likely that a moderate polynomial speed-up is the best one could hope for. Consequently, integrating our approach with QAOA for post-quantum cryptographic analysis is still highly challenging. Recently,  we noticed that the heuristic time complexity of SVP problem was analyzed \cite{prokop2025heuristic}. Furthermore, the parameterized  Closest Vector Problem (CVP) algorithms \cite{priestley2025practically, zalivako2025experimental} are proposed, which are based on the quantum-classical hybrid method\cite{sqif}. The theoretical and practical analysis  offer valuable insights for further performance evaluation of our proposed algorithm. In the future, we plan to explore the LWE problem using alternative methods for solving the Hamiltonian ground state, such as quantum imaginary evolution, quantum Monte Carlo (QMC), quantum annealing, quantum walks and so on.
 Besides the SIS strategy mentioned in this paper, there are other strategies to solve the LWE problem, which are also related to the lattice problem, such as Bounded distance decoding (BDD) strategy. 
Therefore, it is interesting to study which classical strategies are more efficient when combined with the quantum optimization algorithm for the lattice.

\section{methods}

\subsection{The  LWE problem} \label{M1}

\begin{definition}[LWE problem]\label{def1} \cite{Albrecht2015}.
Let $n$ and $q>0$ be integers, $\chi$ be a probability distribution on $\mathbb{Z}$, and $\mathbf{s}$ be a secret vector in $\mathbb{Z}_q^{n}$. Denote  $L_{s,\chi}$ the probability distribution on $\mathbb{Z}_q^{n} \times \mathbb{Z}_q$ obtained as follows: choose $\mathbf{a}$ uniformly at random from $\mathbb{Z}_q^n$, choose $e$ in $\mathbb{Z}_q$ according to $\chi$ and take it modulo $q$, then return
	\begin{equation}\label{eq1}
		(\mathbf{a}, c) = (\mathbf{a}, \braket{\mathbf{a}, \mathbf{s}}_q+e) \in \mathbb{Z}^n_q \times \mathbb{Z}_q,
	\end{equation}
where $\braket{\mathbf{a},\mathbf{s}}_q=(\sum_ia_is_i)\mod q$. 

The LWE-decision problem is the problem of deciding whether pairs $(\mathbf{a},c)\in \mathbb{Z}^n_q\times\mathbb{Z}_q$ are sampled according to $L_{s,\chi}$ or the uniform distribution on $\mathbb{Z}^n_q\times\mathbb{Z}_q$.

The LWE-search problem is to recover $\mathbf{s}$ from $(\mathbf{a},c)=(\mathbf{a},\braket{\mathbf{a,s}}+e)\in \mathbb{Z}^n_q\times\mathbb{Z}_q$.
\end{definition}

Since the LWE-decision problem and LWE-search problem are equivalent \cite{lwe}, therefore, we only deal with the LWE-decision problem in this paper. 

For $m$ pairs $(\mathbf{a}_i, c_i) $ with $e_i$, where $i=1,2,...,m$, 
we define matrix $\mathbf{A}$, vector $\mathbf{e}$, $\mathbf{c}$ as follows
\begin{equation}\label{eq2}
 \mathbf{A} = \left(
 \begin{matrix}
 	\mathbf{a}_1 \\
 	\mathbf{a}_2 \\
 	... \\
 	\mathbf{a}_m 
 \end{matrix}
 \right), \quad
 \mathbf{e} = \left(
 \begin{matrix}
 	e_1 \\
 	e_2 \\
 	... \\
 	e_m
 \end{matrix}
 \right), \quad
 \mathbf{c} = \left(
 \begin{matrix}
 	c_1 \\
 	c_2 \\
 	... \\
 	c_m
 \end{matrix}
 \right).
\end{equation}
Based on Definition \ref{def1}, if the pairs $(\mathbf{a}_i, c_i)$ are obtained according to (\ref{eq1}), we have  the equation $\mathbf{c}=\mathbf{As}+\mathbf{e}\mod q.$ Therefore, the  LWE-decision problem can be described as distinguishing whether $\mathbf{c}=\mathbf{As}+\mathbf{e}\mod q$, or if $\mathbf{c}$ follows a uniform distribution in $\mathbb{Z}_q^m$.   This form of LWE-decision problem is used in our following discussion. 

In this paper, we let $\chi$ be a discrete Gaussian distribution $\mathcal{D}(\mu, \sigma)$ with an average value $\mu=0$ and standard deviation $\sigma$. 
 Someone takes the assumption that the number of samples is unlimited so that we can choose optimal number of samples to minimize the time usage, while others assume that the number of samples is limited. We take the later assumption here.
Given $\mathbf{A} \in \mathbb{Z}^{m \times n}_q$ and $\mathbf{c} \in \mathbb{Z}^m_q$, our goal is to make a decision with a success probability $P > 1/2$. 
By running the decision procedure multiple times, the success probability $P$ can gradually approach $P = 1$. 
We take the assumption that one can achieve this
without using additional samples \cite{Bindel}, thus $m$ samples are enough for the decision.
Consequently, the LWE-decision problem is specified by parameters $n$, $q$, and $\sigma$ and $m$.  Typically, $q$ increase polynomially with $n$, and $\sigma$ increase sublinearly.

\subsection{Solving the LWE-decision problem via SIS approach}\label{M2}

To solve the LWE-decision problem via SIS approach, the essence is that for a vector $\mathbf{v}$ satisfying $\mathbf{A}^T\mathbf{v}=0\mod q$ and a vector $\mathbf{c}$ generated by  $\mathbf{c}=\mathbf{As}+\mathbf{e}\mod q$, we have $I_p=\braket{\mathbf{v}, \mathbf{c}}_q = \mathbf{v}^T(\mathbf{As}+\mathbf{e})\mod q=\mathbf{v}^T\mathbf{e}\mod q = (\sum_i^mv_ie_i)\mod q$. 
Since each independent variable $e_i$ follows Gaussian distribution, $\sum_i^mv_ie_i$  also follows Gaussian distribution. As a comparison, for a vector $\mathbf{c}$ generated uniformly in $\mathbb{Z}_q^m$, $I_p$ follows uniform distribution in $\mathbb{Z}_q$. 
Therefore, we can make the correct decision by generating different vectors $\mathbf{v}$ and observing which  distribution that $I_p=\braket{\mathbf{v},\mathbf{c}}_q$ in $\mathbb{Z}_q$ follows.
  
Heuristically, smaller difference between two probability distributions will lead larger generation times, and we can describe this quantitatively by introducing distinguishing advantage $\epsilon$.
Let $P_1(z)$ and $P_2(z)$ represent the probability functions of Gaussian  distribution and uniform distribution in $\mathbb{Z}_q$ respectively,  where $z=0,1, ... , q-1$. The distinguishing advantage $\epsilon$ of  these two distributions  is defined as $\epsilon=\frac{1}{2}\sum_{z=0}^{q-1}|P_1(z)-P_2(z)|$ \cite{fehr2020sublinear}. Then by $(1/\epsilon)^2$ times sampling, we can distinguish them with success probability close to 1.

The variance of variable $I_{p}' = \sum_i^mv_ie_i$ is equal to $\sum_iv_i^2\sigma^2=||\mathbf{v}||^2\sigma^2$ for $m$ independent variable $e_i$ following Gaussian distribution with variance $\sigma^2$.
Therefore, larger vector length $||\mathbf{v}||$ will increase the variance of Gaussian distribution and make the Gaussian distribution close to uniform distribution in $\mathbb{Z}_q$, thus vanish the distinguishing advantage $\epsilon$.
 To achieve a distinguishing advantage $\epsilon$, 
 the norm of the vector $\mathbf{v}$ should satisfy  \cite{Albrecht2015}
$
    ||\mathbf{v}|| \le (q/\sigma\uppi)\cdot\sqrt{\ln(1/\epsilon)/2}.
$
Therefore, we need to find a sufficiently short  vector $\mathbf{v}$ satisfying both Eq. \ref{length} and $\mathbf{A}^T\mathbf{v}=0\mod q$, which is known as SIS problem, a problem which is considered impossible to solve in polynomial time in classical computing \cite{SIS}.

To solve SIS problem, one can compute a set of vectors $\mathbf{w}_i$ satisfying $\mathbf{A}^T\mathbf{w}_i = 0 \mod q$ by Gaussian elimination method in $\mathbb{Z}_q$, 
and then use them to construct a lattice $\mathcal{L} = \{\sum_iy_i\mathbf{w}_i\mod q | y_i \in \mathbb{Z}_q \}$.
Following this, one can obtain $\mathbf{v}$ by finding short vectors in the lattice $\mathcal{L}$.
 To have an unmodular form of lattice $\mathcal{L}$, we insert $m$ vectors, which are the row vectors of matrix $qI$, into $\mathbf{w}_i$ and compute their reduction basis. The zero-norm vectors obtained in the reduction procedure should be discarded, and we will obtain $m$ independent vectors $\mathbf{b}_i$ with high probability.
Following this line of thought, the LWE-decision problem is  transferred into finding a short enough vector $\mathbf{v}$ in the lattice $\mathcal{L}=\{\sum_i^{m}y_i\mathbf{b}_i|y_i\in\mathbb{Z}_q\}$. 
The procedure of this algorithm is summarized in Algorithm \ref{lwe_alg}.

\begin{algorithm}[H]
	\renewcommand{\algorithmicrequire}{\textbf{Input:}}
	\renewcommand{\algorithmicensure}{\textbf{Output:}}
	\caption{Algorithm for LWE-decision problem via the SIS strategy.}\label{lwe_alg}
	\begin{algorithmic}
		\REQUIRE   $m$ pairs $(\mathbf{a}_i, c_i)$, 		an oracle $O$ that can find the short enough vector for a given lattice.                  
		\ENSURE Decision result.       
		\STATE {\color{blue} \textbf{1.}} Let $\mathbf{A}=(\mathbf{a}_1, \mathbf{a}_2, ... , \mathbf{a}_m)^T$ and $\mathbf{c}=(c_1, c_2, ... , c_m)^T$. Calculate a set of vectors $\{\mathbf{w}_{i\in(m-n)}\}$ that satisfy $\mathbf{A}^T\mathbf{w}_i=0 \mod q$ by the Gaussian elimination method.

	\STATE {\color{blue} \textbf{2.}} Construct the lattice  generated by  $\sum_i^{m-n}y_i\mathbf{w}_i\mod q$ for $y_i\in\mathbb{Z}_q$. 
	Extend the vectors $\mathbf{w}_i$ from $\mathbb{Z}^m_q$ to
	$\mathbb{Z}^m$, and add columns to the lattice matrix, thus change $W = [\mathbf{w}_1, \mathbf{w}_2, ... , \mathbf{w}_{m-n}]$ to
	 $W' = [W | qI]$. Using the LLL algorithm to compute a basis  $\{\mathbf{b}_1, ... ,\mathbf{b}_{m}\}$ from the matrix $W'$.
	
	\STATE {\color{blue} \textbf{3.}} Utilize the oracle $O$ to obtain a sufficiently short  vector $\mathbf{v}$ from lattice $\mathcal{L} = \{\sum_i^{m}y_i\mathbf{b}_i\}$.
	
	\STATE {\color{blue} \textbf{4.}} Calculate the inner product $I_p=\braket{\mathbf{v}, \mathbf{c}}_q$. 
	
	\STATE {\color{blue} \textbf{5.}} Repeat the above procedure to generate enough values of $I_p$. If $I_p$ follows Gaussian distribution in $\mathbb{Z}_q$, output the decision that $\mathbf{c}$ obtained from $\mathbf{c}=\mathbf{As}+\mathbf{e}\mod q$; otherwise,	output the decision that  $\mathbf{c}$ follows the uniform distribution.

	\RETURN Decision result.
	
	\end{algorithmic}
\end{algorithm}

\subsection{LLL algorithm and BKZ algorithm}\label{M3}

In this  paper, we use two classical algorithms,  LLL algorithm and BKZ algorithm, as benchmarks to compare against our proposed method.
 LLL algorithm \cite{LLL} finds an approximate shortest vector $\mathbf{u}$ whose length is $\alpha$ times longer than  the shortest vector in the lattice using polynomial time.
However, the value of $\alpha$ will increase exponentially with the increase of lattice dimension $l$. Therefore, the LLL result will generally not  offer a short enough vector for the  LWE problem. Nevertheless, we can use the LLL algorithm in Step 2 of Algorithm \ref{lwe_alg} to obtain a set of relevant short basis $\{\mathbf{b_i}\}$. Denoting $B$ as the upper bound of the norm of input basis, the time cost of the LLL algorithm can be expressed as $t_L=O(m^{5+\kappa}\log^{2+\kappa}B)$, for every $\kappa>0$ if we employ fast multiplication techniques \cite{LLL,Albrecht2015}.  See `Supplementary Note 1' for the  properties of LLL basis, which is the output of the LLL algorithm.

The BKZ algorithm  \cite{BKZ1987} can find shorter vectors compared to the LLL algorithm, but at the expense of higher complexity. The output quality of the BKZ algorithm is related to the block size $k$ that we choose. Larger $k$ will induce shorter vector with longer calculation time. 
The time cost for the BKZ algorithm is given by 
$t_q=\rho m t_k,$
where $t_k$ refers to the time for calculating SVP of a block with size $k$,
namely, calculating SVP in a lattice with $k$ vectors $\mathbf{b}\in\mathbb{Z}^m$.
There are several methods for calculating SVP. For example, the sieving method takes $t_k=2^{0.2972k}$ operations and memory heuristically \cite{laarhoven2015faster}, while
the enumeration method \cite{yasuda2021survey} takes poly$(k)$ memory and $2^{O(k^2)}$ operations \cite{fincke1985improved}.
  $\rho$ represents the number of  times that we need to call the SVP oracle. An empirical equation of $\rho$ is 
$\rho=m^2/k^2\cdot \log m$\cite{Albrecht2015, hanrot2011analyzing}. 

Since larger block size $k$ will induce shorter vector and longer calculation time, we should confirm how to choose $k$ for a given LWE-decision problem.
After calculation, we find that 
\begin{equation}\label{eq15}
	k={m}/{\log \big(q^{1-n/m}/\sigma\uppi\cdot \sqrt{\ln(1/\epsilon)/2}\big)}
\end{equation}
 is sufficient to achieve the distinguishing advantage $\epsilon$.
(See `Supplementary  Note 3' for the derivation.) 

 Since the time complexity of BKZ algorithm increases exponentially with the block size $k$,  
it is interesting to see which number of samples $m$ will lead to the minimum $k$ for given $n,q,\sigma,\epsilon$. We let $\partial k/\partial m = 0$ 
to derive
\begin{equation}
	m = \frac{2n}{1+ \log(\frac{1}{\sigma\uppi }\sqrt{\frac{1}{2}\ln\frac{1}{\epsilon}})/\log q}.
\end{equation}
Therefore, $m\approx 2n$ is a suitable choice of the number of samples, and we take this relation in our simulation. Take $q=\text{poly}(n)$, $m\propto n$ into Eq. \ref{eq15}, we can find that $k$ is proportional to $m$ and $n$.

\subsection{Strategy for initializing parameters}

 The periods of $E(\mathbf{\gamma}, \mathbf{\beta})$ regarding the parameters $\beta_i$ and $\gamma_i$ are $\uppi$ and $2\uppi$ respectively.   In the simulation, we observe that fact that $E(\mathbf{\gamma}, \mathbf{\beta})$ changes rapidly and reaches the maximum value and minimum value when $\gamma_k$ is small. When $\gamma_k$ becomes large, barren plateau appears, which means that the change of $E(\mathbf{\gamma}, \mathbf{\beta})$ is quite small when $\gamma_k$ varies. 

 We give a heuristic explanation why the phenomenon occurs. We write a state $\ket{\psi}$ as $\ket{\psi}=\sum_jc_j\ket{j}$, where $\ket{j}$ are the eigenstates of the Hamiltonian.  The operator $e^{-i\gamma_k H}$ induces a phase rotation $e^{-i\gamma_k E_j}$ in front of each eigenstate $\ket{j}$ in the state $\ket{\phi}$. In the following, we let $\phi_j=\gamma_kE_j+t\cdot 2\uppi$, where the integer $t$ is chosen to restrict $\phi_j$ to the range $[-\uppi,\uppi)$.
 When $|\gamma_k|$ is  small,  the values of  $ \gamma_k E_j$ are in the range $[-\uppi,\uppi)$ for all of $j$. Consequently, increasing $\gamma_k$ will deterministically lead to an increase in  $\phi_j$  for every $j$, resulting in significant changes to $\ket{\psi}$ and $\braket{\psi|O|\psi}$. However, when $|\gamma_k|E_j$ become much larger than $\uppi$ for different $j$, it becomes quite randomly that whether $\phi_j$ will increase or decrease by increasing $\gamma_k$, and in statistics,  there may no obvious influence of the change of the parameters. Therefore, the change of $\gamma_k$ will not influence the expectation value of the Hamiltonian significantly.

Following this, we can restrict the value of $\gamma_k$ into the range  $[-2\uppi/E_a,2\uppi/E_a)$ to avoid the possible barren plateau phenomenon, where $E_a$ is the averaged value of $E_j$. On the other hand,  QAOA is inspired from the quantum annealing algorithm, and the parameter $\gamma$, $\beta$ in the QAOA are corresponding to $t/T$ and $(1-t/T)$ respectively in the quantum annealing. Therefore, we can gradually decrease the value of $\beta_k$ and increase the value of $\gamma_k$ when $k$ increases. Following this, we can set the initial parameters as
     \begin{equation}
     	\gamma_k = -\frac{2\uppi}{E_a}\cdot\frac{k}{p+1},\qquad
     	\beta_k = \frac{\uppi}{4}\cdot\frac{p-k+1}{p}.
     \end{equation}
The Ising  Hamiltonian  can be written as $H = \sum_{i,j}a_{ij}\sigma^z_i\sigma^z_j+\sum_ib_i\sigma^z_i+c$, and roughly,  we can take $E_a\approx c$.

\section*{acknowledgements}
S.W. acknowledges the Beijing Nova Program under Grants No. 20230484345 and 20240484609;  We also acknowledges the National Natural Science Foundation of China under grant No. 62471046.

\section*{Author contributions}
 M. Z.  and S. W.  developed the theoretical framework. S. W. and G.L. L. supervised the work.   W. Y. and B. Y. contribute to the  proof of the theorem. M. Z, J. Z, P.J. C., Q. L., H. Z., and M. W. contribute to the simulation and the experiment. All authors contributed in the preparation of the manuscript.

\section*{Competing interests}

The authors declare no competing interests.

\section*{Data availability} 

The data that support the findings of this work are provided in Supplementary Data 1.

\section*{Code availability}

The code that supports the findings of this work is available from the authors
on reasonable request.

\renewcommand{\bibsection}{} 
\section*{References} 
\bibliographystyle{naturemag}
\citestyle{nature}
\bibliography{HAWI.bib}

\clearpage

\appendix

\section*{ Supplementary Information for "Quantum-Classical Hybrid Algorithm for Solving the Learning-With-Errors Problem  on NISQ Devices"}

\section*{Supplementary Note 1: The LLL Reduction Basis }\label{append_lll}

By applying Schmidt orthogonalization to the LLL basis\cite{LLL}
$\{\mathbf{b}_1, \mathbf{b}_2, ... , \mathbf{b}_l\}$, we obtain the transformed basis vectors as
\begin{equation}
	\mathbf{\tilde{b}}_i=\mathbf{b}_i-\sum_{j=1}^{i-1}\mu_{j,i}\mathbf{\tilde{b}}_j, \quad \mu_{j,i}=\frac{\braket{\mathbf{b}_i, \mathbf{\tilde{b}}_j}}{\braket{\mathbf{\tilde{b}}_j,\mathbf{\tilde{b}}_j}}.
\end{equation}
We denote $\{\mathbf{\tilde{b}}_1, \mathbf{\tilde{b}}_2, ... , \mathbf{\tilde{b}}_l\}$ as the Schmidt orthogonal basis.    It is worthy to  note that this orthogonal basis is not necessarily in the lattice.  The LLL basis satisfies the following conditions: 

\textbf{(1)} For all $ j<i$, it holds that
\begin{equation}\label{eqa2}
	|\mu_{i,j}|\le1/2. 
\end{equation}
When expanding the basis $\{\mathbf{b}_1, \mathbf{b}_2, ... , \mathbf{b}_l\}$
using the orthogonal basis  $\{\mathbf{\tilde{b}}_1, \mathbf{\tilde{b}}_2, ... , \mathbf{\tilde{b}}_n\}$,
and organizing the coefficients into a matrix, inequality (\ref{eqa2}) can be represented as:

\begin{equation}\label{eq3}
\begin{aligned}
	\left( \mathbf{b}_1, \mathbf{b}_2, ... , \mathbf{b}_l \right) &=
	\left(
	\begin{matrix}
		|\tilde{\mathbf{b}}_1| & \mu_{1,2}|\tilde{\mathbf{b}}_1| & \mu_{1,3}|\tilde{\mathbf{b}}_1| & ... & \mu_{1,l}|\tilde{\mathbf{b}}_1| \\
		& |\tilde{\mathbf{b}}_2| & \mu_{2,3}|\tilde{\mathbf{b}} _2|&  ... & \mu_{2,l}|\tilde{\mathbf{b}}_2| \\
		&&& ... \\
		& & &  &|\tilde{\mathbf{b}}_l|
	\end{matrix}
	\right)\\
	&=
	\left(
	\begin{matrix}
		|\tilde{\mathbf{b}}_1| & <\frac{1}{2}|\tilde{\mathbf{b}}_1| & <\frac{1}{2}|\tilde{\mathbf{b}}_1| & ... & <\frac{1}{2}|\tilde{\mathbf{b}}_1| \\
		& |\tilde{\mathbf{b}}_2| & <\frac{1}{2}|\tilde{\mathbf{b}} _2|&  ... & <\frac{1}{2}|\tilde{\mathbf{b}}_2| \\
		&&& ... \\
		& &  &  & |\tilde{\mathbf{b}}_l|
	\end{matrix}
	\right).
\end{aligned}
\end{equation}

\textbf{(2)} The adjacent orthogonal basis vectors satisfy the following equation:
\begin{equation}\label{eqa4}
	||\mathbf{\tilde{b}}_{i+1}||^2\ge(\delta-\mu_{i,i+1}^2)||\mathbf{\tilde{b}}_i||^2.
\end{equation} 
Combining equations (\ref{eqa2}) and (\ref{eqa4}), we have
\begin{equation}
	|| \mathbf{\tilde{b}}_{i+1} ||^2\ge (\delta-1/4)||\mathbf{\tilde{b}}_i||^2.
\end{equation}
Consequently, we can conclude that $||\tilde{\mathbf{b}}_{i}||\ge (\delta-1/4)^{\frac{i-1}{2}}||\tilde{\mathbf{b}}_1||$.

\section*{Supplementary Note 2: Upper Bound of Qubit Number} \label{apendix_bound}

To find shorter vectors, we begin with the LLL basis $\{\mathbf{b}_1, \mathbf{b}_2, ... , \mathbf{b}_l\}$. Let $\mathbf{v}$ represents the vector in the lattice, namely $\mathbf{v}=\sum_ic_i\mathbf{b}_i$, and define $R=||\mathbf{b}_1||$. 
We aim to analyze the possible values of the coefficients $c_i$
for a vector $\mathbf{v}_s$
that satisfies $||\mathbf{v}_s||<R$.

(1)
From equation (\ref{eq3}), it is evident that the component of $\mathbf{v}_s$ on the last orthogonal basis $\tilde{\mathbf{b}}_l$ is contributed solely by $\mathbf{b}_l$. Therefore, the coefficient on  $\mathbf{b}_l$ must satisfy $|c_l|<R/||\mathbf{\tilde{b}}_l||$, otherwise only 
 $\mathbf{v}_s\cdot \mathbf{\tilde{b}}_l/|| \mathbf{\tilde{b}}_l ||$, the component of $\mathbf{v}_s$, will be larger than $R$. Consequently,
\begin{equation}
	|c_l|<\frac{R}{||\mathbf{\tilde{b}}_l||}\le \frac{R}{(\delta-\frac{1}{4})^{\frac{l-1}{2}}R}=(\delta-\frac{1}{4})^{\frac{1-l}{2}}.
\end{equation}
Denote this upper bound of $|c_l|$ as $\gamma_{l}$, hence we have $\gamma_l=(\delta-\frac{1}{4})^{\frac{1-l}{2}}$.

(2) The component of $\mathbf{v}_s$ on the orthogonal basis is contributed by $\mathbf{b}_{l-1}$ and $\mathbf{b}_l$. The properties of the LLL basis indicate that the projection of $\mathbf{b}_l$ onto $\tilde{\mathbf{b}}_{l-1}$, namely  $\mu_{l-1,l}||\tilde{\mathbf{b}}_{l-1}||$, doesn't exceed $|| \tilde{\mathbf{b}}_{l-1} ||/2$. Therefore, for all vectors $\mathbf{v}_s$ satisfying $||\mathbf{v}_s||<R$, their component $\mathbf{b}_l$ will offer no more than length $\gamma_l \cdot||\tilde{\mathbf{b}}_{l-1} ||/2$ on the orthogonal basis $\tilde{\mathbf{b}}_{l-1}$. Consequently, if we let the coefficient $c_{l-1}$ satisfy:
\begin{equation}
\begin{aligned}
 	|c_{l-1}|_{\text{max}}&= \frac{R+(\delta-\frac{1}{4})^{\frac{1-l}{2}}\cdot\frac{1}{2}|| \tilde{\mathbf{b}}_{l-1}||}{||\tilde{\mathbf{b}}_{l-1}||} \\
	& =\frac{R}{||\tilde{\mathbf{b}}_{l-1}||}+\frac{1}{2}(\delta-\frac{1}{4})^{\frac{1-l}{2}} \\
	&\le
	(\delta-\frac{1}{4})^{\frac{2-l}{2}} + \frac{1}{2}(\delta-\frac{1}{4})^{\frac{1-l}{2}}, 
\end{aligned}
\end{equation}
then all possible vectors will be covered. Therefore,  $\gamma_{l-1}=(\delta-\frac{1}{4})^{\frac{2-l}{2}} + \frac{1}{2}(\delta-\frac{1}{4})^{\frac{1-l}{2}} $.

(3) Similarly, the component of $\mathbf{v}_s$ on the orthogonal basis $\tilde{\mathbf{b}}_{l-2}$ is provided by $\mathbf{b}_{l-2}$, $\mathbf{b}_{l-1}$ and $\mathbf{b}_l$. 
For all vectors $\mathbf{v}_s$ satisfying
 $||\mathbf{v}_s||<R$, their
 component $\mathbf{b}_l$ offers no more than $\gamma_l\cdot||\tilde{\mathbf{b}}_{l-2} ||/2$ on the orthogonal basis
 $\tilde{\mathbf{b}}_{l-2}$, 
 while their component $\mathbf{b}_{l-1}$ offers no more than $\gamma_{n-1}\cdot||\tilde{\mathbf{b}}_{l-2} ||/2$ on the orthogonal basis $\tilde{\mathbf{b}}_{l-2}$. Therefore, 
\begin{equation}
	\begin{aligned}
		|c_{l-2}|_{\text{max}} =& \frac{R+\gamma_l\cdot\frac{1}{2}||\tilde{\mathbf{b}}_{l-2} ||+\gamma_{l-1}\cdot\frac{1}{2}||\tilde{\mathbf{b}}_{l-2} ||}{||\tilde{\mathbf{b} }_{l-2}||}\\
		&=\frac{R}{||\tilde{\mathbf{b}}_{l-2} ||}+\frac{\gamma_l+\gamma_{l-1}}{2}\\
		&\le (\delta-\frac{1}{4})^{\frac{3-l}{2}}+\frac{\gamma_l+\gamma_{l-1}}{2}.
	\end{aligned}
\end{equation}

(4) Following this way, we can obtain the recursive relation of $\gamma_i$ as
\begin{equation}
	\gamma_{l-i} = (\delta-\frac{1}{4})^{\frac{1+i-l}{2}}+\frac{1}{2}\sum_{k=0}^{i-1}\gamma_{l-k}
\end{equation}
with initial condition
$\gamma_l=(\delta-\frac{1}{4})^{\frac{1-l}{2}}
$.
Let
\begin{equation}
	\gamma_{l-i} =\alpha_i (\delta -1/4)^{\frac{1-l}{2}},
\end{equation}
then  $\alpha_i$ satisfies the following recursive relation:
\begin{equation}
	\alpha_i =(\delta-\frac{1}{4})^{\frac{i}{2}}+\frac{1}{2}\sum_{k=0}^{i-1}\alpha_k
\end{equation}
with initial condition
$
	\alpha_0=1
$, 
which is independent with $l$. The analytical expression of $\alpha_i$ is difficult to obtain, but the first few $\alpha$ can be easily computed as follows:
\begin{equation}
	\begin{aligned}
		&\alpha_0=1 \\
		& \alpha_1=\sqrt{\delta-\frac{1}{4}}+\frac{1}{2}\\
		& \alpha_2=(\delta-\frac{1}{4})+\frac{1}{2}\sqrt{\delta-\frac{1}{4}}+\frac{1}{2^2}+\frac{1}{2}\\
		& ...
	\end{aligned}
\end{equation}

As an example, let $\delta=3/4$, then $\gamma_{l-i}=2^{\frac{l-1}{2}}\cdot\alpha_i$ and  $\alpha_i=2^{-\frac{i}{2}}+\frac{1}{2}\sum_{k=0}^{i-1}\alpha_k$. 
For $i \in[1, 2000]$, which is the range we usually focus on in post-quantum cryptography, we can find that $\log_2\alpha_i$ is consistently smaller than $i$, so there is $\alpha_i < 2^i$. Therefore,
\begin{equation}
	\gamma_{l-i}=2^{\frac{l-1}{2}}\cdot\alpha_i < 2^{\frac{l-1}{2}+i}.
\end{equation}

In the quantum algorithm, the coefficients are encoded by Pauli matrices. $j$ qubits can represent $2^j$ coefficients, 
therefore, we can use $\lceil\log_2( 2\gamma_{l-i})\rceil=\lceil\frac{l+1}{2}\rceil+i$ qubits to encode $(l-i)$-th LLL reduction basis. 

Consequently, the number of total qubits we need is
\begin{equation}
	S  = l\cdot \lceil\frac{l+1}{2}\rceil +\sum_{i=0}^{l-1} i =l\cdot \lceil\frac{l+1}{2}\rceil + \frac{l(l-1)}{2} 
\end{equation}

Therefore,
\begin{equation}
		S = 
	\left\{
	\begin{aligned}
		l^2, \quad l\ \text{is odd.} \\
		l(l+1), \quad l\ \text{is even.}
	\end{aligned}
	\right.   
	\le l(l+1). 
\end{equation}

\section*{Supplementary Note 3: Appropriate block size $k$ of BKZ algorithm}\label{appendc}

Denote $\text{vol}(\mathcal{L})$ as the determinant of a lattice $\mathcal{L}$. For an output vector $\mathbf{b}$, the Hermite factor $\delta_0$ is defined as
\begin{equation}
	||\mathbf{b}||=\delta_0^m\cdot\text{vol}(\mathcal{L})^{1/m}.
\end{equation}
Since the Gaussian heuristic states that the shortest vector $\mathbf{v}_0$ in the lattice satisfies $||\mathbf{v}_0||=\sqrt{m/(2\pi e)}\text{vol}(\mathcal{L})^{1/m}$, the Hermite factor describes the difference between the shortest vector $\mathbf{v}_0$ and the approximate shortest vector $\mathbf{b}$.

Heuristically, the relation between $\delta_0$ and block size $k$ follows
$
	\delta_0 = 2^{1/k}
$.\cite{Albrecht2015}
For the LWE-decision problem, we have
$
	\text{vol}(\mathcal{L}) = q^n
$.\cite{Albrecht2015}
Therefore, we have
\begin{equation}\label{C2}
||\mathbf{b}|| = 2^{m/k}\cdot q^{n/m}.
\end{equation}
The relation between vector norm and distinguishing advantage $\epsilon$ follows
\begin{equation}\label{C3}
	||\mathbf{b}|| = (q/\sigma\pi)\cdot\sqrt{\ln(1/\epsilon)/2}.
\end{equation}
Combine Eq. (\ref{C2}) and (\ref{C3}), we have
\begin{equation}
	k =\frac{m}{\log \big(q^{1-n/m}/\sigma\pi\cdot \sqrt{\ln(1/\epsilon)/2}\big)}.
\end{equation}
If we choose $\epsilon=1/e^2$, we have
\begin{equation}
	k=\frac{m}{\log \big(q^{1-n/m}/\sigma\pi\big)}.
\end{equation}
Thus the time complexity is
\begin{equation}
	T=\frac{e^4m^3}{k^2}\log m\cdot 2^{0.2972k}=e^4m\log m\cdot f2^{0.2972m/f},
\end{equation}
where $f=f(q,n,m,\sigma)=\log(q^{1-n/m}/\sigma\pi)$.

\section*{Supplementary Note 4: Analysis of qubit number}

From the proof in Supplementary Note 3, we can find that if we use shorter vectors as the basis to encode the Hamiltonian, the required qubits are expected to be fewer.
Therefore, due to the loose theoretical restriction condition for the LLL basis, it seems impossible to decrease the theoretical bound of the qubit number if we directly use the LLL basis to encode the Hamiltonian. M. Albrecht et al. \cite{Albrecht2023} gives a more tight bound of qubit number $O(m\log m)$ for SVP by using the property of Hermite Korkine-Zolotarev (HKZ) basis \cite{kannan1983improved}, which follows a much tight restriction condition compared to LLL basis.
However, because finding the HKZ basis is even more computationally complex than solving the SVP, it is not feasible to directly use the HKZ basis for encoding the Hamiltonian. The algorithm  proposed by M. Albrecht et al. \cite{Albrecht2023} treats this matter wisely by running the quantum algorithm  multiple times, which leads to  more time usage of the algorithm.

\section*{Supplementary Note 5: Numerical details}

In the simulation and experiment, we focus on the LWE-decision problem with a  small $n$ for demonstration. In such cases, the LLL reduction basis is sometimes already sufficiently short for the decision. 
 Therefore, the impact of finding the shortest vector may not be significantly.

We introduce the simulation process of Fig. 2(a) in the main text in detail. The parameter is set as $n=18$, $m=36$, $q=331$, $\sigma=3$.
To begin with, we generate a vector $\mathbf{s}\in \mathbb{Z}^{n}_{q}$ as the secret vector,  $M=9\times 10^5$
matrices $\mathbf{A}\in \mathbb{Z}^{m\times n}_{q}$ as the samples, and $M$ error vectors $\mathbf{e}\sim\mathcal{D}(0, \sigma)$. 
$M$ vectors $\mathbf{c}$ are generated by  $\mathbf{c}=\mathbf{As}+\mathbf{e}\mod q$, while 
another $M$ vectors $\mathbf{c}$ are generated randomly in $\mathbb{Z}^m_q$.
After $2M$ pairs $(\mathbf{A},\mathbf{c})$ are generated, we could start the solving process. Firstly, we transform the problem into the short vector problem, and  
calculate the shortest vector $\mathbf{v}_0$ in the lattice,  which could be obtained if we use $O(m^2)$ qubits and successfully find the first excited state of the Hamiltonian in Eq. (4) in the main text.
In the calculating process, we observe that there are $R$=743385 instances where the SVP in the lattice is shorter than $\mathbf{b}_0$ in LLL basis. We use the shortest vectors and LLL vectors to calculate the inner product $I_p$ respectively.
Since the variance $\sigma$ is relatively small,  $I_p$ is more likely appear close to 0 or $q$ for those instances sampled according to $L_{s,\chi}$. In this case, we can set a bound $I_b$ for each calculation. If $I_p\in[0,I_b]\cup[q-I_b,q]$, we deduce that the samples are generated by $L_{s,\chi}$. Otherwise, if $I_p\in[I_b, q-I_b]$, we deduce that $\mathbf{c}$ is randomly generated  in $\mathbb{Z}^{m}_{q}$. By repeating the calculation, the success probability of the decision will be close to 1.  
By optimizing $p_b$, we obtain a maximum probability of correct decision $P_{\max}=0.581$ at $p_b=81$ for each calculation. 
Specifically,  the number of correct decisions is $m_1=523259$, in comparison to $m_2=510371$ if we directly use LLL basis for calculation. 
 The fluctuation of results caused by the randomness of samples matches the standard variance for $R$ instances of random walk,  which is $\sigma_r = \sqrt{2R}$. Since $m_1-m_2>10\sigma_r=10\sqrt{2R}$, there is a certain improvement of the success probability after vector reduction.

In Fig. 2(b) in the main text, we randomly generated $M=2000$ Matrices $\mathbf{A}$ for the simulation. In Fig. 2(c) in the main text, for problem size $n\ge12$, we let $M=5000$, and for $n<12$, since the proportion that LLL basis doesn't include the shortest vector in the lattice is small, we let $M=50000$ to generate enough instances to calculate success probability for different qubit number. In Fig. 2(d) in the main text, 
    we select instances where one qubit per LLL basis is sufficient to identify the shortest vector  for the convenience of simulation.
     We simulate 5 LWE instances for each point.  The error bar includes the range from $P_{\min}$ to $P_{\max}$.

\section*{Supplementary Note 6:  Experimental details }

\begin{figure*}[h]
	\centering
	\includegraphics[width=0.55\textwidth]{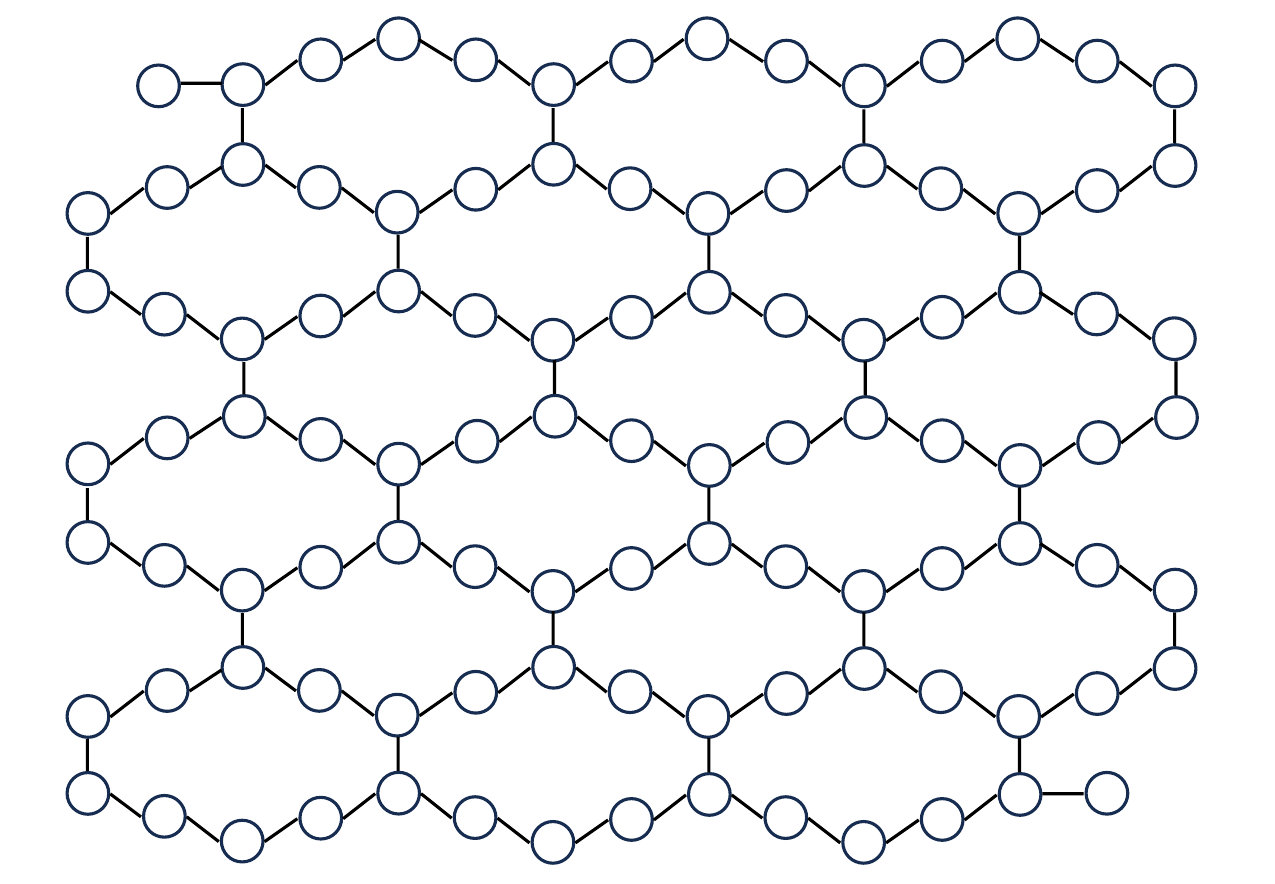}
	\caption{The architecture of ibm\_kyoto\cite{kyoto, roy2024simulating}}\label{fig_s0}
\end{figure*}

\begin{figure*}[h]
	\centering
	\includegraphics[width=1\textwidth]{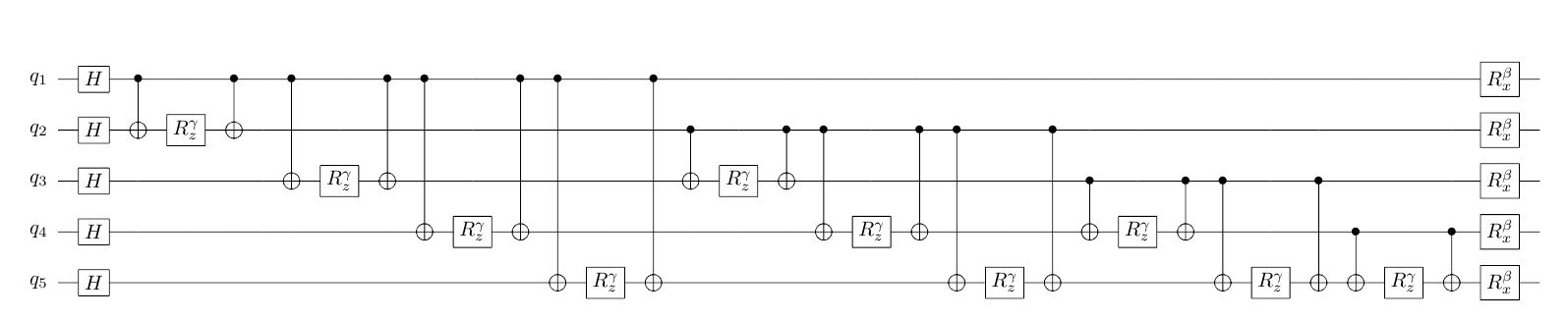}
	\caption{The parameterized circuit for the experiment. The number of qubits is 5, and the layer of QAOA is 1.}\label{fig_s1}
\end{figure*}

\renewcommand\arraystretch{1.5}
\begin{table*}[h]
 \caption{Calibration data of the `ibm\_kyoto'}\label{t1}
 \resizebox{0.6\textwidth}{!}{
  \begin{tabular}{cccccc}
   \toprule[1.5pt]
   \quad 	property \quad & \quad  $T_1(\mu s)$ \quad  & \quad  $T_2(\mu s)$ \quad  & 
    \quad  ECR error \quad  & \quad  Readout error \quad  \\
	\hline
	\quad  Median \quad  &\quad  223.49\quad  & \quad 118.92\quad 	 &\quad  8.12e-3\quad  & \quad 1.53e-2\quad  \\
   \bottomrule[1pt]
   \end{tabular} }
\end{table*}

\begin{table}[h]
\caption{The expectation values of the Hamiltonian for each iteration in the experiment.}\label{t2}
\resizebox{0.95\textwidth}{!}{
	\begin{tabular}{cccccccccc}
    \toprule[1.5pt]
    Repeatation&iteration 1&iteration 2 &iteration 3&iteration 4&iteration 5 & iteration 6 & iteration 7 & iteration 8 & iteration 9
    \\
    \midrule[1pt]
    1& 24.927 & 23.030 & 18.881 & 20.692 & 20.806 & 19.921 & 15.615 & 14.575 & 12.563 \\
    2& 25.189 & 26.207 & 24.278 & 24.540 & 23.130 & 20.523 & 18.271 & 15.762 & 15.365 \\
    3& 22.213 & 22.774 & 23.186 & 23.690 & 19.007 & 17.711 & 17.043 & 16.722 & 15.433\\
    4& 25.914 & 24.036 & 24.054 & 20.330 & 20.384 & 19.915 & 15.448 & 15.337 & 14.622\\
    \midrule[1pt]
    Average & 24.561 &  24.012 & 22.600 & 22.313 &  20.832 &  19.517 & 16.594 & 15.600 & 14.496 \\
    Standard variance & 1.403 & 1.352 & 2.185 & 1.831 & 1.484 & 1.072 & 1.150 & 0.775 & 1.160\\
    \bottomrule[1.5pt]
\end{tabular}}
	\label{tab:sherbrooke}
\end{table}

There is a little trick when we use classical optimization method to minimize $E(\beta,\gamma)$. 
For our Hamiltonian $H=\sum_{i,j}a_{i,j}\sigma_i^z\sigma_j^z+\sum_ib_i\sigma^z_i$, the coefficient $b_{i,j}$ is much larger than 1, making that 
\begin{equation}
	E(\beta, \gamma)=\braket{\phi_0|e^{i\beta\sum_i\sigma_x}\cdot e^{-i\gamma H} \cdot H\cdot e^{-i\beta\sum_i\sigma_x}\cdot e^{-i\gamma H}|\phi}
\end{equation}
change more rapidly with $\gamma$ than $\beta$. 
To avoid this problem, we multiply a scale factor $1/s_{\text{a}}$ $(s_{\text{a}}\gg1)$ for $H$ in the unitary operation $e^{-i\gamma H}$ in the optimization process. (Or equivalently, we can decrease the learning rate in the $\gamma$ direction from $r$ to $r/s_{\text{a}}$.)

We use gradient descent method for optimization. For the $i$-th iteration step, we update the parameter by $\mathbf{\theta}_i=\mathbf{\theta}_{i-1}-r\nabla E(\mathbf{\theta}_{i-1})$, where $r$ is the learning rate. The gradient $\nabla E(\mathbf{\theta})$ at point $\mathbf{\theta_0}=(\gamma_0, \beta_0)$ is calculated by
	\begin{equation}
		\begin{aligned}
			\nabla E({\mathbf{\theta_0}})=&\Big(\frac{E(\gamma_0+\Delta,\beta_0)-E(\gamma_0,\beta_0)}{\Delta}, \\ &\qquad\frac{E(\gamma_0,\beta_0+\Delta)-E(\gamma_0,\beta_0)}{\Delta}\Big).
		\end{aligned}
	\end{equation}
	In the experiment, we choose $\Delta=0.05$ and  learning rate $r=0.06$.

 We use the quantum device `ibm\_kyoto' in IBM quantum platform to run our algorithm. The topology of this hardware for the experiments is shown in Supplementary Figure \ref{fig_s0}. The quantum circuit is compiled to ECR,  RZ, SX, X, ID gates before the implementation. The calibration data  of the `ibm\_kyoto' \cite{kyoto, roy2024simulating} is listed in Supplementary  Table \ref{t1}. The exact circuit is shown in Figure \ref{fig_s1}. We take 8192 shots (namely 8192 QAOA samples) to get one expectation value. The expectation value in the experiment is listed in  Table \ref{t2}.

\section*{Supplementary Note 7: Heuristic Estimation of Time Complexity}

\begin{figure*}
\centering
	\subfloat[\label{fig:3a}]{\includegraphics[width=0.45\textwidth]{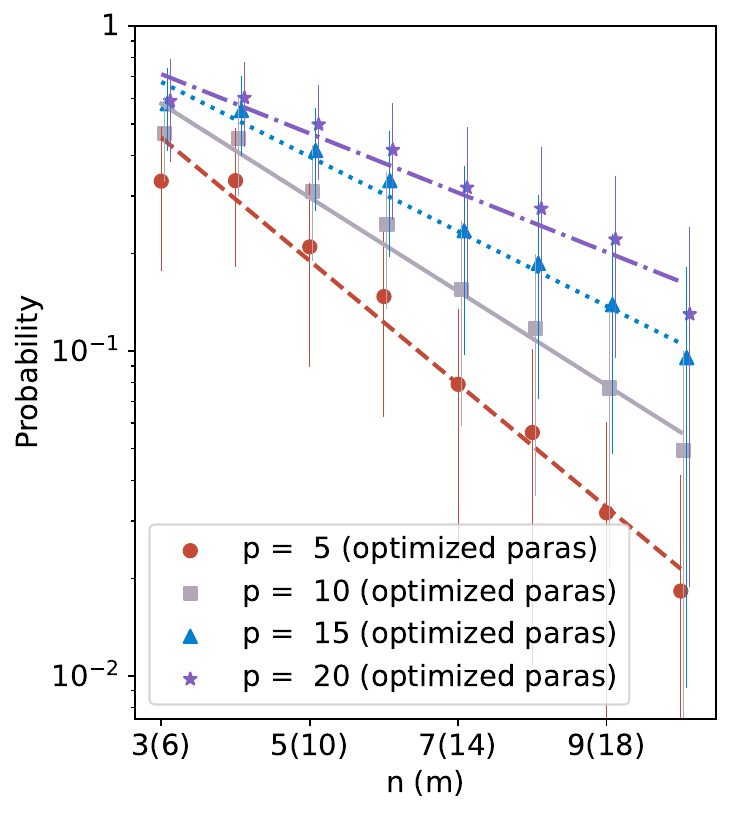}}
	\subfloat[\label{fig:3b}]{\includegraphics[width=0.45\textwidth]{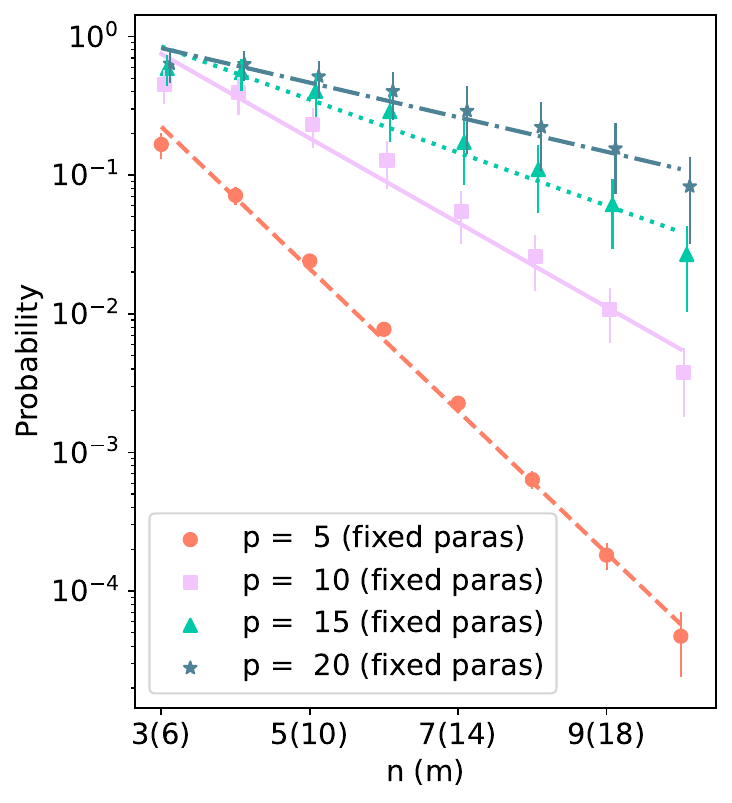}}
	\caption{The success probability of QAOA as a function of problem size. (a) Success probabilities with optimized parameters obtained using the gradient descent method. (b) Success probabilities with parameters fixed according to Eq. (11) in the main text. The error bar on each data point $(n_i, P_i)$ extends from $(n_i, P_i - \sigma_i)$ to $(n_i, P_i + \sigma_i)$, where $\sigma_i =\sqrt{ \frac{1}{R} \sum_{j=1}^R(P_{ij}- \frac{1}{R} \sum_{k=1}^RP_{ik})^2}$ for $R$  simulation results $P_{ij}$, $j=1,2, ... , R$.
	}\label{fit}
\end{figure*}

\renewcommand\arraystretch{1.5}
\begin{table*}[h]
\caption{Heuristic Estimation of Time Complexity with Optimized Parameters}\label{t3}
\resizebox{1.0\textwidth}{!}{
  \begin{tabular}{cccccc}
   \toprule[1.5pt]
   \quad  Layer \quad & \quad  $p=5$ \quad  & \quad  $p=10$ \quad  &  
    \quad  $p=15$ \quad  & \quad  $p=20$ \quad  \\
	\hline
	\quad  Fitted Curve (Expontial) \quad &\quad  $\exp(-0.17m - 0.087)$ \quad  & \quad $\exp(-0.14m + 0.10)$\quad   
	 &\quad  $\exp(-0.12m + 0.12)$\quad  & \quad $\exp(-0.11m + 0.067)$\quad  \\
   \bottomrule[1pt]
   \end{tabular} }
\end{table*}

\renewcommand\arraystretch{1.5}
\begin{table*}[h]
\caption{ Heuristic Estimation of Time Complexity with Fixed Parameters}\label{t4}
\resizebox{1.0\textwidth}{!}{
  \begin{tabular}{cccccc}
   \toprule[1.5pt]
   \quad  Layer \quad & \quad  $p=5$ \quad  & \quad  $p=10$ \quad  &  
    \quad  $p=15$ \quad  & \quad  $p=20$ \quad  \\
	\hline
	\quad  Fitted Curve of Time Complexity \quad &\quad  $\exp(-0.57m + 1.3)$ \quad  & \quad $\exp(-0.32m + 1.2)$\quad   
	 &\quad  $\exp(-0.19m + 0.70)$\quad  & \quad $\exp(-0.12m + 0.34)$\quad  \\
   \bottomrule[1pt]
   \end{tabular} }
\end{table*}

Here, we provide a heuristic estimation of how the time complexity of our algorithm increases with problem size by simulating the algorithm at small scales. We fix the number of layers in the QAOA circuit at \( p = 5, 10, 15, 20 \) and increase the problem size to observe the trends. The results of our simulations are presented in Figure \ref{fit}. 

In Figure \ref{fig:3a}, we initialize the parameters according to Eq. (11) in the main text and then optimize them using the gradient descent method to obtain the optimized quantum state \(\ket{\psi_{\text{op}}} = U \ket{\psi_{\text{ini}}}\). The success probability is then calculated using \( p = \braket{\psi_{\text{targ}} | \psi_{\text{op}}} \), where \(\ket{\psi_{\text{targ}}}\) represents the ground state of the Hamiltonian. For each data point in the figure, we randomly generate 50 instances and fit the results using an exponential function of the form \( T = \exp(am + b) \). The fitted functions are summarized in Table \ref{t3}.

Inspired by the work of A. Montanaro et al. \cite{PRXQuantum.5.030348}, Figure \ref{fig:3b} displays another approach where we do not optimize the parameters, resulting in the success probability \( p' = \braket{\psi_{\text{targ}} | \psi_{\text{ini}}} \). By fixing the parameters, we avoid potential optimization challenges, such as barren plateaus, that can arise as the problem size increases. This approach provides a more reliable, albeit potentially pessimistic, estimation of the time complexity. The fitted functions for this approach are presented in Table \ref{t4}.

It is important to note that, since the problem sizes considered in our simulation are relatively small, the fitted curves should be regarded as the heuristic estimation of the time complexity of our algorithm. The time complexity for larger problem sizes may not necessarily conform to the form \( T \propto \exp(am + b) \).

\end{document}